\documentclass[11pt,twoside]{article}
\usepackage{asp2010}
\usepackage{natbib}

\resetcounters

\begin{document}

\title{Multifluid magnetohydrodynamic turbulence in weakly ionised
	astrophysical plasmas}
\author{Turlough P.\ Downes$^{1,2,3}$
\affil{$^1$School of Mathematical Sciences, Dublin City University,
	Glasnevin, Dublin 9, Ireland.}
\affil{$^2$School of Cosmic Physics, Dublin Institute for Advanced
	Studies, 31 Fitzwilliam Place, Dublin 2, Ireland.}
\affil{$^3$National Centre for Plasma Science and Technology, Dublin
	City University, Glasnevin, Dublin 9, Ireland.}}

\begin{abstract}
Initial results from simulations of 4-fluid MHD turbulence in molecular 
clouds are presented.  The species included in the simulations are ions, 
electrons, negatively charged dust grains and neutrals.  The results indicate 
that, on length scales of a few tenths of a parsec, multifluid effects have a
significant impact on the properties of the turbulence.  In particular,
the power spectra of the velocity and magnetic fields are significantly
softened, while the PDF of the densities of the charged and neutral
fluids are appreciably different.  Indeed, the magnetic field strength
displays much less spatial structure on all lengthscales up to 1\,pc
than in the ideal MHD case.  The assumptions of ideal MHD
therefore appear to be inadequate for simulating turbulence in molecular 
clouds at these length scales.
\end{abstract}

\section{Introduction}
It is generally believed that molecular clouds are turbulent (see the
reviews of \citealt{mac04,elm04}).  Observationally \citep[e.g.][]{larson81} 
this turbulence appears to be supersonic, with RMS Mach numbers of anything up
to 10, and either super-Alfv\'enic or trans-Alfv\'enic.  The amplitude of this turbulence makes it 
likely to be an important ingredient in our understanding both of the dynamics of molecular clouds 
and of the process of star formation \citep{elmegreen93} and hence developing an
understanding of this phenomenon is of considerable interest.

Many authors have addressed the issue of MHD turbulence in the context
of molecular clouds using both the ideal MHD approximation
\citep[e.g.][]{mac98, mac99, ost01, glo07, lem09} and, more recently, various 
flavours of non-ideal MHD \citep{ois06, li08, kud08, dow09, dow11}.
Ambipolar diffusion has been found to cause greater temporal variability
in the turbulence statistics.  Clearly of lesser
significance, the Hall effect is in principle capable of inducing topological 
changes in the magnetic field which are quite distinct to any influence caused 
by ambipolar diffusion.

The non-ideal MHD effects caused by multifluid physics occur on relatively 
small length scales (fractions of a parsec).  
However, since three dimensional MHD turbulence involves the transfer of 
energy from the energy injection scale to ever smaller scales until the 
dissipation length scale of the system is reached, it is virtually inevitable 
that multifluid effects will have an impact on some part of the energy 
cascade.  If the energy transport is non-local in $k$-space then these
effects may influence the entire energy cascade.

\citet{dow09, dow11} performed simulations of decaying,
non-ideal MHD and multifluid MHD, molecular cloud turbulence incorporating 
parallel resistivity, the Hall effect and ambipolar diffusion.  They found 
that the Hall effect has surprisingly little impact on the behaviour of the 
turbulence: it does not affect the energy decay at all and has very limited 
impact on the power spectra of any of the dynamical variables, with the 
exception of the magnetic field at very short lengthscales. 

In this paper we use the HYDRA code \citep{osu06, osu07} to 
investigate driven, isothermal, multifluid MHD turbulence in a system
consisting of 1 neutral and 3 charged fluids.  The aim of this work is to
determine the influence of multifluid effects on the behaviour of driven 
turbulence in molecular clouds.  An added benefit of our properly multifluid 
approach is that we can self-consistently deduce the behaviour of the charged
species and, in principle, make links with observations such as those of
\citet{lh08}.

\section{The model}
\label{sec:model}

Molecular clouds are, to a great extent, weakly ionised systems.  This
allows us to make several simplifying assumptions, as follows:
the bulk flow velocity {\em is} the neutral velocity; the majority of 
collisions experienced by each charged species occur with neutrals; the 
charged species' inertia is unimportant and, finally, the charged species' 
pressure gradient is unimportant.  Making these assumptions allows us to 
derive a relationship a generalised Ohm's law \citep[e.g.][]{falle03, cio02}, 
and this allows us to avoid having to calculate the electric field explicitly 
from the charge distribution.

The resulting weakly ionised, multifluid MHD equations are a nonlinear
system of PDEs \citep[see][]{cio02, falle03, osu07} and, in general, must be 
tackled using simulations.  The method used by HYDRA to integrate these 
equations has been elucidated previously in the literature 
\citep[e.g.][]{osu06, osu07} and 
we refer the reader to these papers for a more in-depth discussion.  
The method is explicit and scales extremely well on massively parallel 
systems, displaying strong scaling from 8192 up to 294,912 cores with 73\% 
efficiency on the JUGENE BlueGene/P system.  

\section{Numerical set-up}
\label{sec:initial-conditions}

We run our turbulence simulations in a cube of side 1\,pc with periodic
boundary conditions on all faces.  The density and magnetic field are
initially uniform with values of $10^6$\,cm$^{-3}$ and 1\,mG respectively.  The
(isothermal) sound-speed is chosen to be $3.3\times10^4$\,cm\,s$^{-1}$ or,
equivalently, the temperature is 30\,K.  The four fluids in the system are 
initially at rest and the properties of the charged species are chosen
to approximate those of dust grains, ions and electrons with the dust
grains making up 1\% of the mass density of the neutrals.  We require
charge neutrality and that the resistivities in the generalised Ohm's
Law be as described in \citet{war99}.  The grid used is uniform and made up 
of $N_{\rm g}^3$ grid points with $N_{\rm g}$ points in each of the $x$, $y$ 
and $z$ directions.  We emphasise here that this work is 
directed at understanding the nature of multifluid MHD turbulence.  While a 
physical system with the properties just described may well be subject to 
dynamically important self-gravity, it seems reasonable that the optimal way 
of approaching understanding the behaviour of molecular clouds is to attempt 
to understand the turbulence itself before including the effect of
self-gravity which will, undoubtedly, significantly impact the behaviour
of the system.

To drive turbulence we add velocity increments to the neutral
velocity at each time-step.  The incremental velocity field, $\mathbf{\delta u}$, 
is defined in a manner similar to \cite{lem09}.  Each component of 
$\mathbf{\delta u}$ is generated from a set of waves with wave numbers, 
$3 \leq k \leq 4$ where $k = |\mathbf{k}|$.  The amplitudes of these waves are 
drawn 
from a Gaussian random distribution with mean 1.0 and deviation 0.33 while the 
phases of the waves are drawn from a uniform distribution between 0 and 
$2 \pi$.  A new realisation of the incremental velocity field is
generated after $10^{-3}\,t_{\rm c}$ has elapsed since the previous
realisation was generated.

The rate of injection of energy is the same for all simulations in this work 
and is given by $\frac{\dot{E}}{\rho_0 L^2 c_{\rm s}^3} = 300$ where $\rho_0$ 
is the initial (uniform) density, $L$ is the length of side of the 
computational domain and $c_{\rm s}$ is the sound speed.  In our simulations 
this energy injection rate yields steady state turbulence with an RMS Mach 
number of around 5 and an Alfv\'enic Mach number (with respect to the 
unperturbed magnetic field) of around 1.2.

The nomenclature used here to refer to the simulations is defined as
follows: each simulation is referred to as either mf-xxx or mhd-xxx with
``mf'' standing for multifluid and ``mhd'' standing for ideal MHD.  The
``xxx'' gives the value of $N_{\rm g}$.

\section{Resolution study}

\begin{figure}
\plottwo{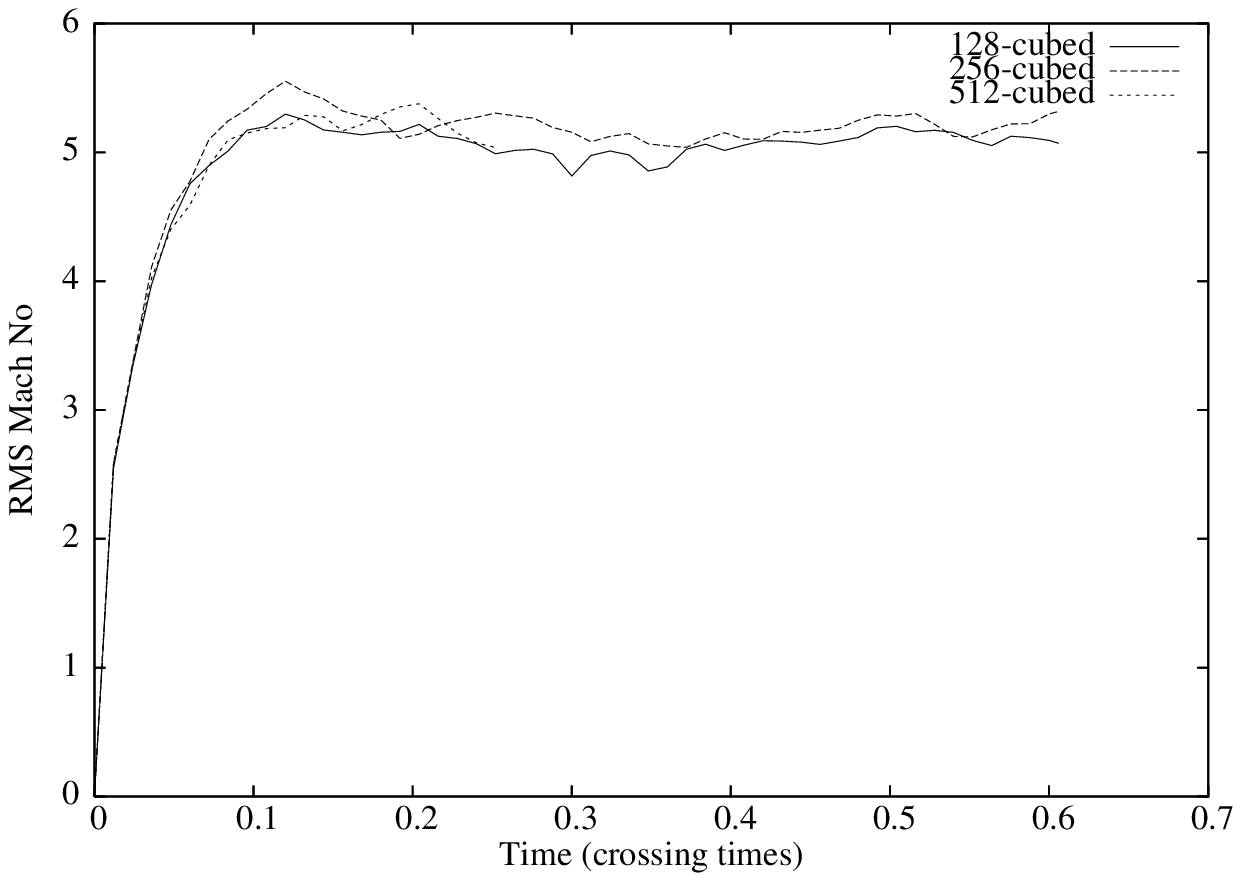}{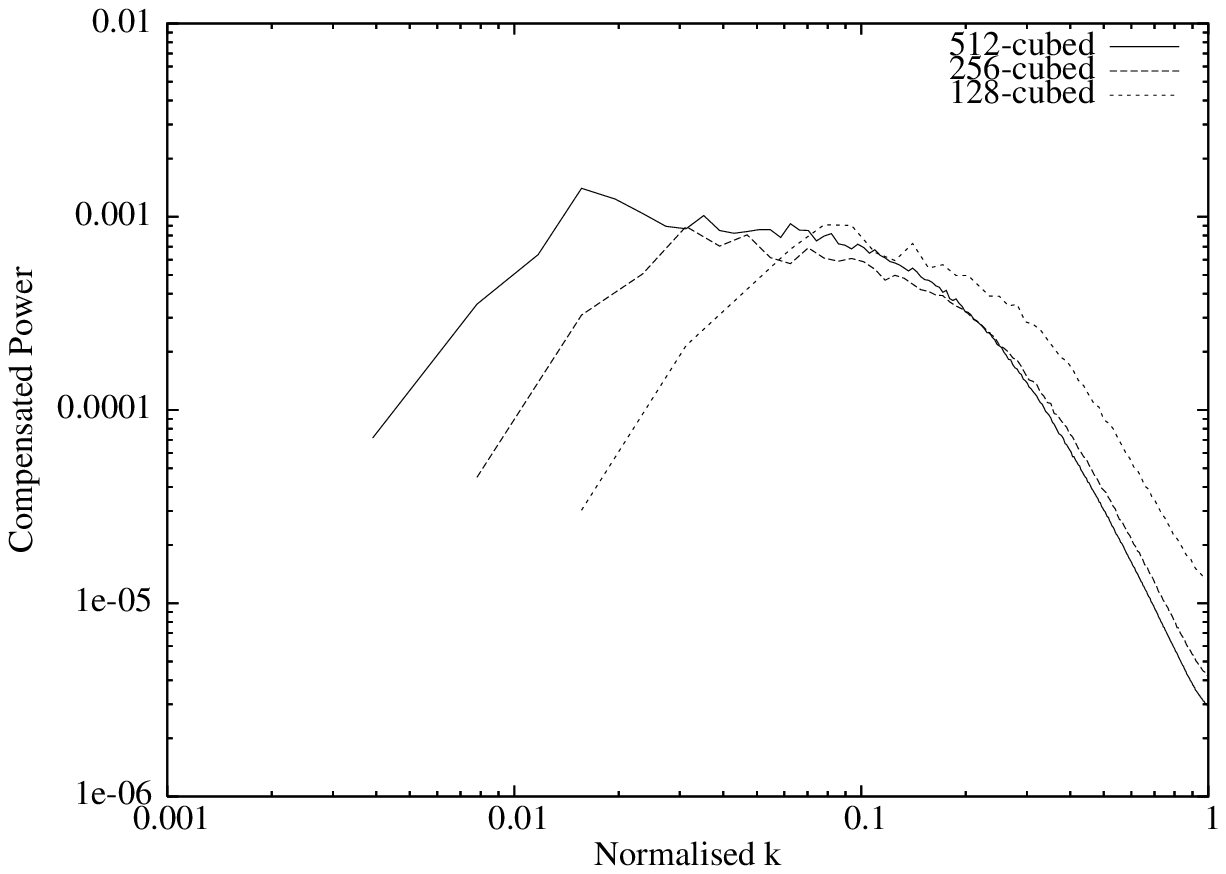}
\caption{Evolution of the RMS Mach number with time (left panel) and the
time-averaged compensated neutral velocity power spectra (right panel)
for each simulation in the resolution study.}
\label{downes_fig:res-study}
\end{figure}

Figure \ref{downes_fig:res-study} contains the Mach number as a function of 
time and the compensated power spectra of the neutral velocity for three simulations 
(mf-128, mf-256 and mf-512) in our resolution study.  While there are 
differences between mf-128 and mf-256 it is clear that the differences between 
mf-256 and mf-512 are minor, indicating that both simulations resolve 
the multifluid physics occurring in this system reasonably well.

\section{Results}

In this section we analyse some aspects of the turbulence generated in
our simulated systems.  We compare the properties of the mf-512
simulation with that of mhd-512 in order to gain a sense of the
differences introduced by multifluid effects at the 1\,pc level.
The left panel of Figure \ref{downes_fig:res-study} indicates that for
$t \geq 0.12$\,$t_{\rm c}$ the turbulence has reached a quasi-steady steady 
state.  Any time-averaged quantities discussed in this section are averaged 
over times after $t = 0.12$\,$t_{\rm c}$.

\subsection{Energy dissipation}
\label{downes_sec:dissipation}

\begin{figure}
\centering
\plottwo{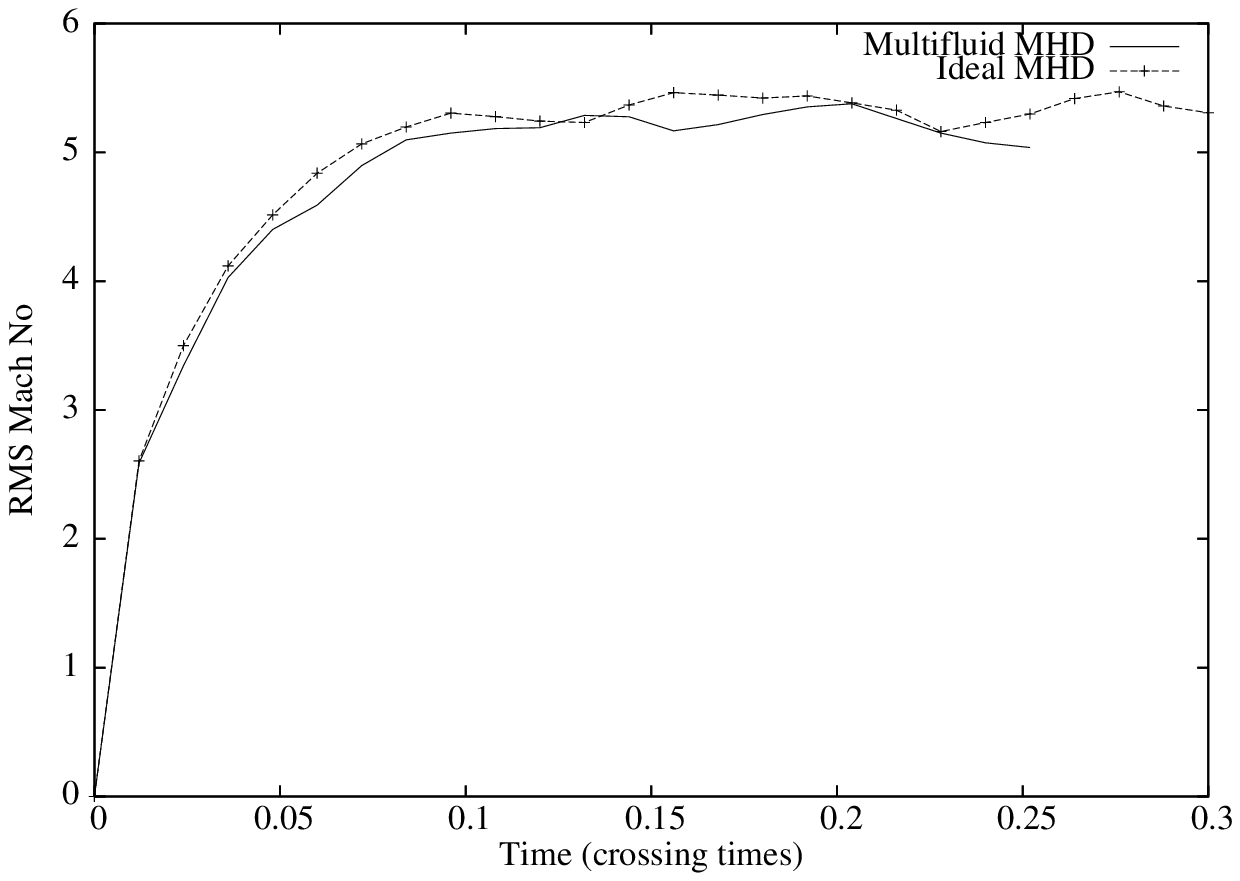}{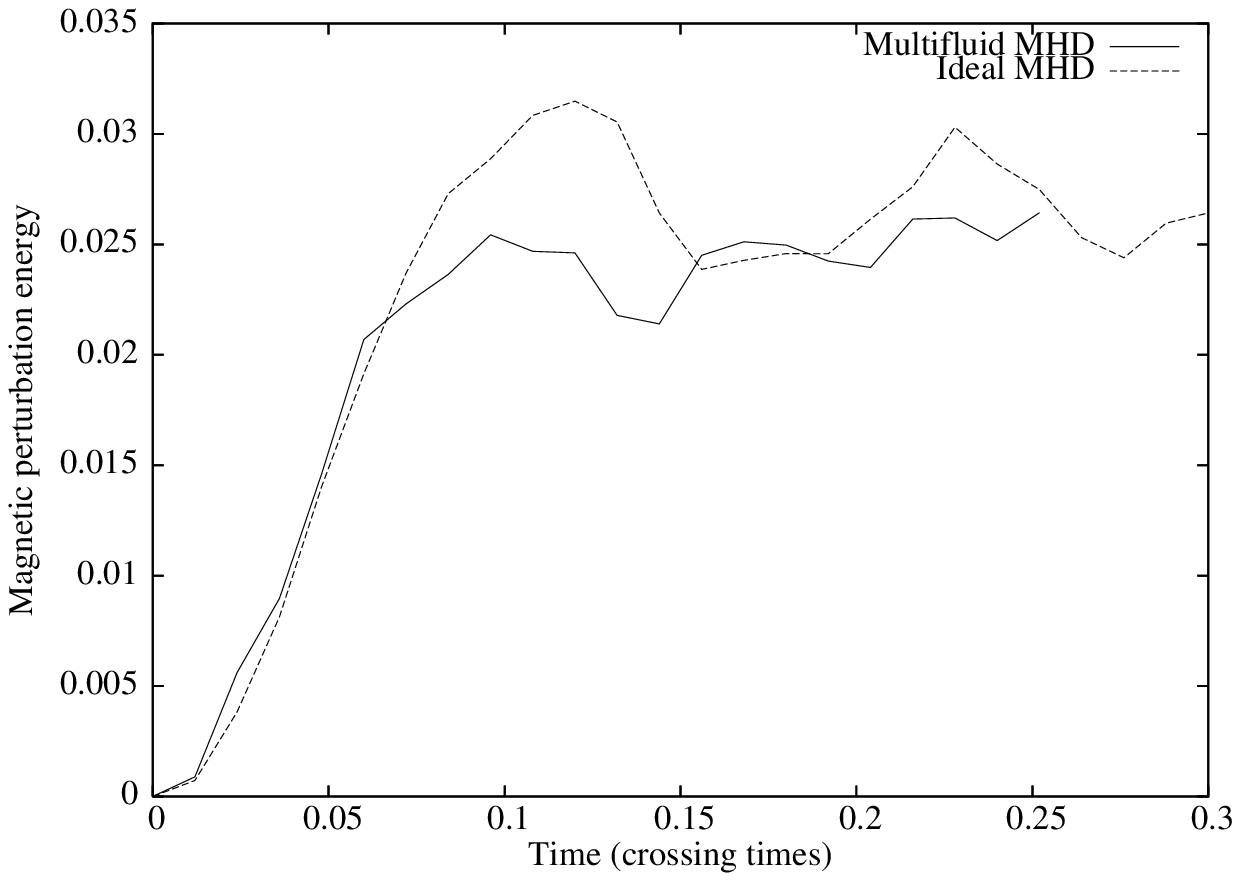}
\caption{Plot of the RMS Mach number (left panel) and the perturbed
	magnetic energy (right panel) as a function of time for mf-512 and 
	mhd-512.}
\label{downes_fig:mf-mhd-rms}
\end{figure}

Figure \ref{downes_fig:mf-mhd-rms} contains plots of the behaviour of
the mean, mass-weighted RMS Mach number \citep[see][]{lem09, dow11} and
the perturbed magnetic energy as functions of time.  It is clear
that the Mach number reached in each case is very similar and in fact
the difference in time-averaged RMS Mach number after $t=0.12$ is less
than 1\%.  Therefore, at these length scales and with this numerical 
resolution, we do not detect any difference in the energy dissipation rate 
due to multifluid effects.  It is worth noting, however, that the
perturbed magnetic energy in mf-512 is lower than that in mhd-512
(Figure \ref{downes_fig:mf-mhd-rms}, right panel) which
indicates that the energy in the turbulence is less weighted toward the
magnetic field in a multifluid MHD system than in an ideal MHD one.

The work of \cite{dow09, dow11} implies that the energy dissipation rate
is higher in multifluid MHD systems.  Their simulations were carried out in a 
cube of length 0.2\,pc and the decay rate of the turbulence is undoubtedly 
higher in their multifluid simulation than in the ideal case.  While
this point is worthy of further study, it may be the case that 
the difference in the length scales, which determines the physical resolution 
of the simulations, may be partly responsible for the apparent 
discrepancy in the results presented here and those of \cite{dow09, dow11}.  

\subsection{Power spectra}

We now turn to the power spectra of the various quantities of interest
in these simulations.  The nature of these spectra will give us some idea of the influence
of the multifluid effects on the turbulence.  We time-average all the power spectra
discussed here over the time-interval $[0.12\,t_{\rm c}:0.22\,t_{\rm c}]$. 

\begin{figure}
\centering
\plottwo{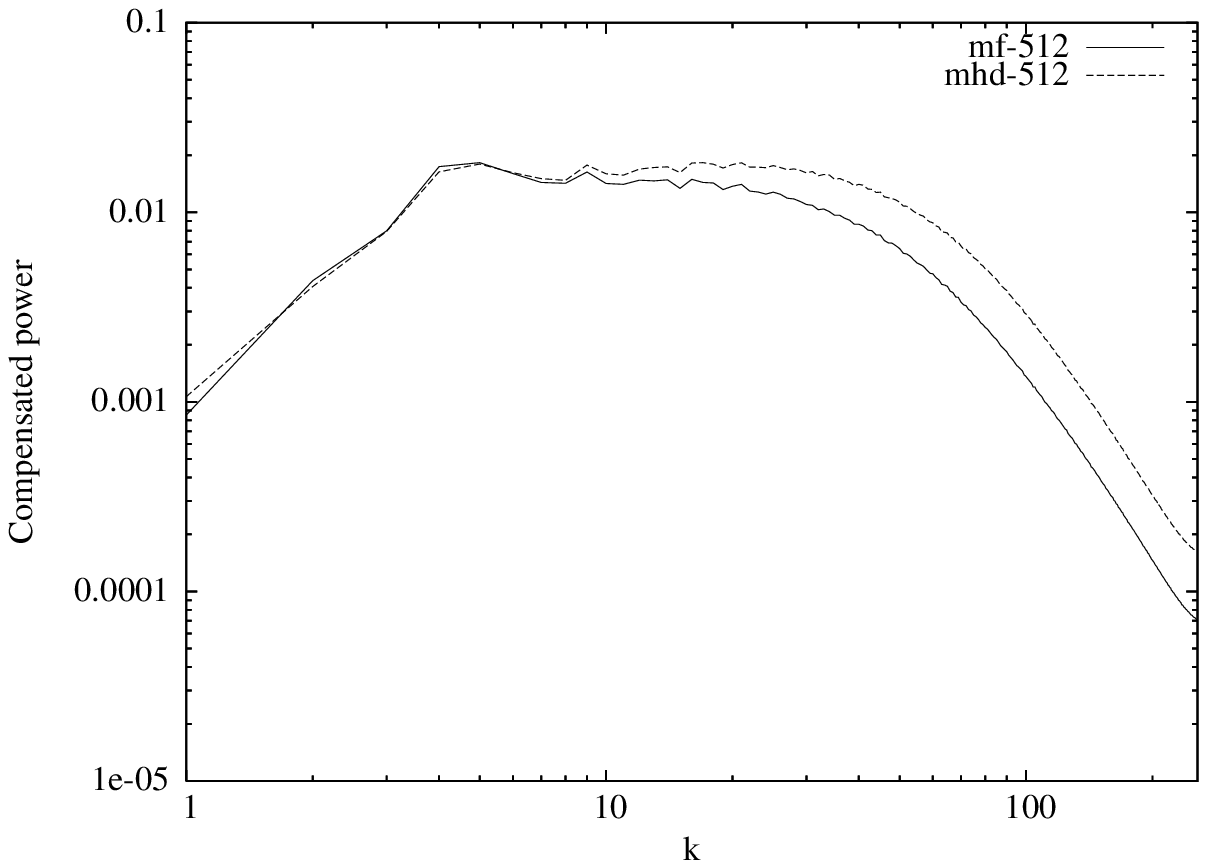}{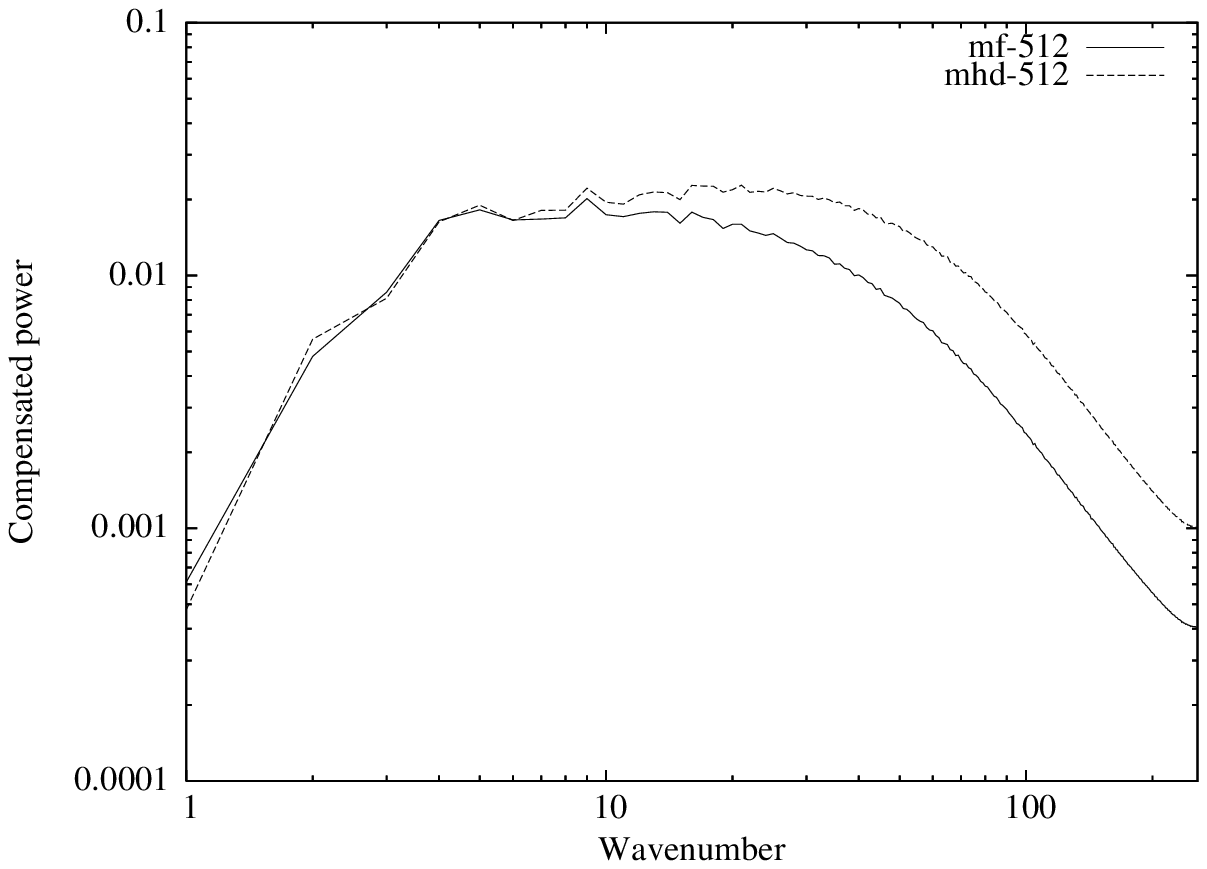}
\caption{Time-averaged, compensated power spectra of the velocity field (left panel) and the
	magnetic field strength (right panel) for mhd-512 and mf-512.  The velocity spectra are
		compensated with $k^{1.5}$ while the magnetic field spectra are compensated with $k^2$.}
\label{downes_fig:power-spec-mf-mhd}
\end{figure}

Figure \ref{downes_fig:power-spec-mf-mhd}  contains plots of the velocity and magnetic field
strength power spectra for mhd-512 and mf-512.  It is clear that the velocity field in mhd-512 has less 
structure at length scales below about 0.05\,pc than that of mf-512.  The difference between
the two systems is more dramatic in the power spectra of the magnetic field strength where
simulation mf-512 displays considerably less structure at virtually all length scales below the
driving scale.  This is, perhaps, what we would expect given the fact that the multifluid effects
manifest themselves most directly in the induction equation.

\begin{figure}
\centering
\plottwo{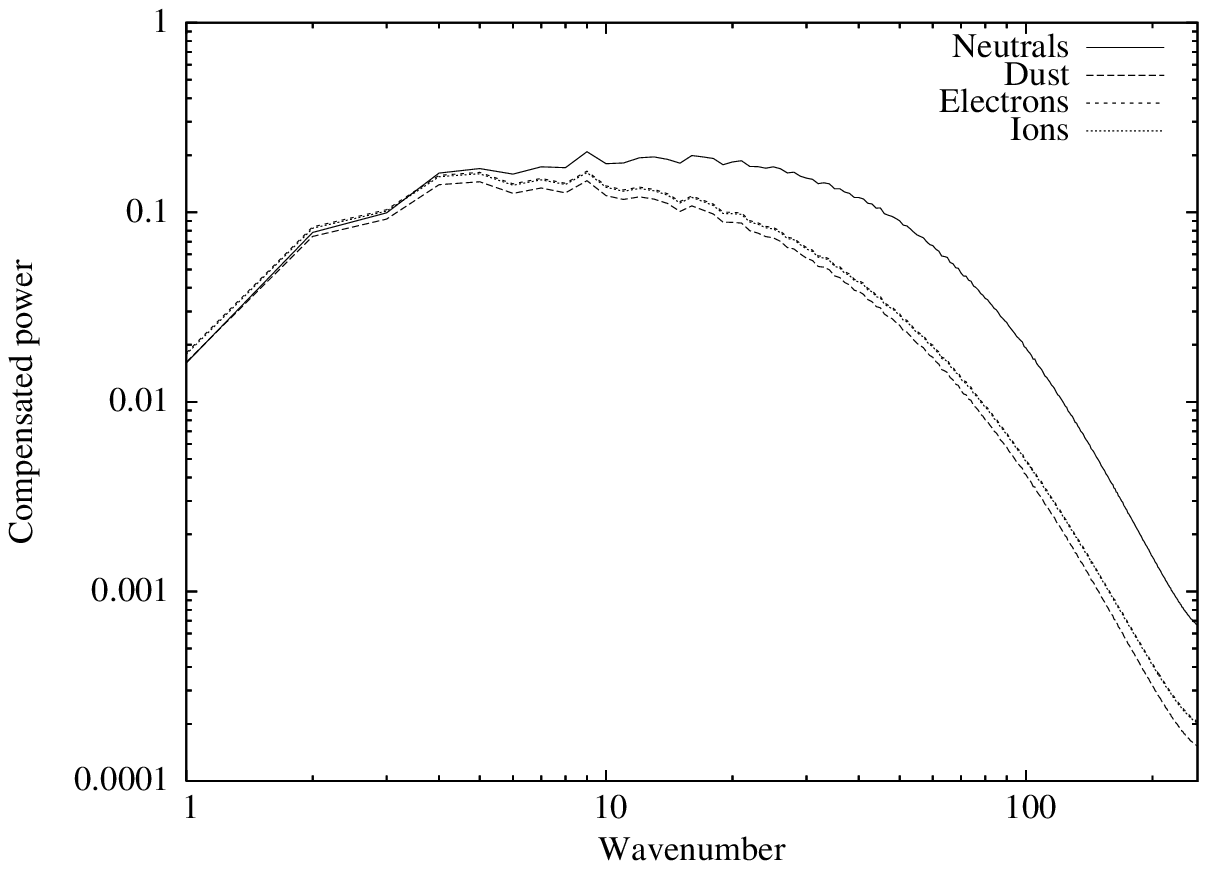}{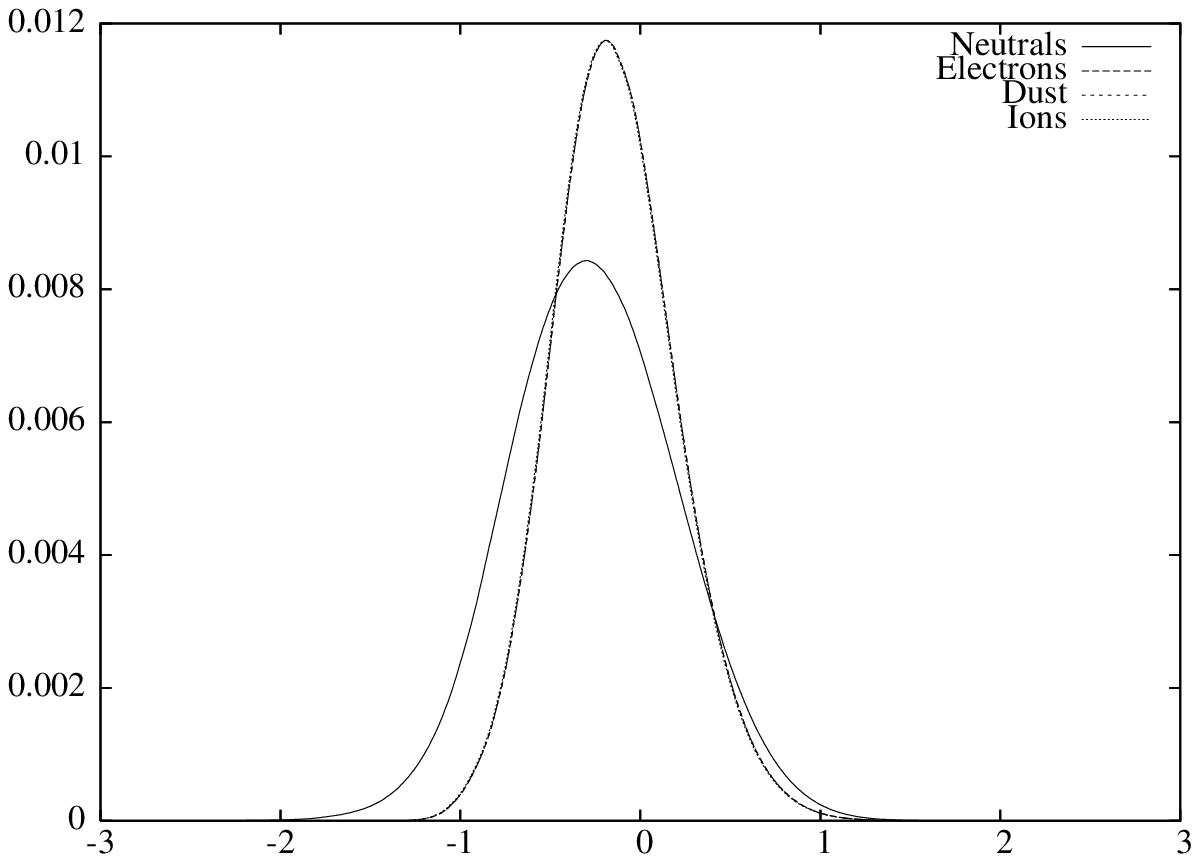}
\caption{Time-averaged, compensated power spectra of the densities of the various fluids in
	mf-512 (left panel) and the probability distribution functions of the logarithm of their densities 
		(right panel).}
\label{downes_fig:power-spec-rho-mf}
\end{figure}

The left panel of figure \ref{downes_fig:power-spec-rho-mf} contains plots of the power spectra of the 
densities for all of the fluids in mf-512.  At these length scales the electrons and ions have almost 
identical power spectra while the neutral fluid has considerably more power at all lengthscales
below $\sim0.1\,$pc.  These differences are due to the fact that the electrons and ions are more
closely tied to the dynamics of the magnetic field than are the neutrals.  The magnetic field, as a
result of ambipolar diffusion, has less power at shorter length scales in particular (see Figure
		\ref{downes_fig:power-spec-mf-mhd}), and 
hence the species closely tied to it also have less fine-scale structure.  Interestingly the density
PDFs of the fluids in mf-512 (right panel of figure \ref{downes_fig:power-spec-rho-mf}) indicate that, 
though all fluids are distributed in an approximately lognormal way, the distributions of the charged 
species are considerably more peaked and narrower.  This is another manifestation of the effect just 
noted: since the magnetic field is somewhat decoupled from the bulk (neutral) flow it is, in particular, 
not compressed or rarefied by the turbulent flow as much as the neutral flow.  This means that the 
species tied to the magnetic field will not be compressed or rarefied as much as the neutrals and hence 
the PDF of their densities will have a lower standard deviation than that of the neutrals.

\section{Conclusions}

In this work we present 4-fluid MHD simulations of turbulence in molecular clouds, neglecting
self-gravity.  The aim is to investigate the impact of multifluid MHD effects on the turbulence
itself while the inclusion of self-gravity will be the subject of future work.  The results indicate 
that, on the lengthscales at which the simulations were run ($\sim 1\,$pc) multifluid effects have an
appreciable effect on both the velocity and density power spectra, and there is a significant
difference between the PDFs of the neutral and charged species.

This all implies that, even on fairly large length scales, multifluid effects are dynamically
important in molecular cloud turbulence.

\acknowledgements 
This material is based upon works supported by the Science Foundation Ireland
under Grant No.  07/RFP/PHYF586.  We thank the DEISA Consortium
(www.deisa.eu), funded through the EU FP7 project RI-222919, for support
within the DEISA Extreme Computing Initiative.  The authors also wish to 
acknowledge the SFI/HEA Irish Centre for High-End Computing (ICHEC) for the 
provision of computational facilities and support.

\bibliography{tdownes.bib}
\end{document}